\PassOptionsToPackage{table}{xcolor}  
\PassOptionsToPackage{colorlinks=true, linkcolor=blue, citecolor=red}{hyperref}
\documentclass[10pt]{wlscirep}
\usepackage[utf8]{inputenc}
\usepackage[T1]{fontenc}

\usepackage{amsmath}
\usepackage{graphicx}
\usepackage{amssymb} 
\usepackage{physics}
\usepackage{epstopdf}
\usepackage{color}

\usepackage{enumitem}

\begin{document}

\title{Parameter and hidden-state inference in mean-field models from partial observations of finite-size neural networks}

\author[1,*]{Irmantas Ratas}
\author[1]{Kestutis Pyragas}

\affil{Center for Physical Sciences and Technology, LT-10257 Vilnius, Lithuania}
\affil[*]{irmantas.ratas@ftmc.lt}

\begin{abstract}
We study large but finite neural networks that, in the thermodynamic limit, admit an exact low-dimensional mean-field description. We assume that the governing mean-field equations describing macroscopic quantities such as the mean firing rate or mean membrane potential are known, while their parameters are not. Moreover, only a single scalar macroscopic observable from the finite network is assumed to be measurable. Using time-series data of this observable, we infer the unknown parameters of the mean-field equations and reconstruct the dynamics of unobserved (hidden) macroscopic variables. Parameter estimation is carried out using the differential evolution algorithm. To remove the dependence of the loss function on the unknown initial conditions of the hidden variables, we synchronize the mean-field model with the finite network throughout the optimization process. We demonstrate the methodology on two networks of quadratic integrate-and-fire neurons: one exhibiting periodic collective oscillations and another displaying chaotic collective dynamics. In both cases, the parameters are recovered with relative errors below $1\%$ for network sizes exceeding 1000 neurons.

\end{abstract}


\maketitle

\section{\label{sec1} Introduction}

Real neural networks are highly complex dynamical systems composed of billions of interacting neurons. In neuroscience, understanding and predicting the macroscopic behavior of such systems is crucial, as it underlies essential physiological functions of neural networks, and their dysfunctions may lead to neurological diseases. Direct modeling of neural networks at the microscopic level using detailed models of individual neurons and synapses is extremely computationally challenging. To overcome this complexity, low-dimensional neural mass models, which mimic the coarse-grained activity of large neuronal populations, have been developed long ago~\cite{Wilson1972,Destexhe2009}. These models are represented by closed low-dimensional systems of differential equations that describe the collective activity of neural populations in terms of average quantities such as the mean membrane potential or mean firing rate~\cite{Wilson1972,Destexhe2009,Jansen1995,Jirsa1996,Spiegler2010}.
However, such models are phenomenological in nature: they do not explicitly account for microscopic synchronization mechanisms and therefore have a limited scope of application~\cite{Devalle17}. 

Recently, a new class of low-dimensional mean-field descriptions, known as next-generation neural population models (NGNPM)~\cite{Coombes2023}, has been developed. In contrast to phenomenological approaches, these models are derived directly from the underlying microscopic dynamics and become exact in the thermodynamic limit of infinitely many neurons. Their success relies on the existence of a low-dimensional invariant manifold onto which the trajectories of the high-dimensional microscopic system collapse~\cite{Ott2008,Montbrio2015}. For networks of quadratic integrate-and-fire (QIF) neurons, an explicit analytical representation of this manifold enables an exact reduction of the microscopic equations to a closed mean-field system. Although the NGNPM theory is valid only for a certain class of neurons, it can be expected that a low-dimensional description of macroscopic dynamics should be possible for other classes of neurons as well. One possible route toward constructing such descriptions is the use of modern data-driven techniques, for example the sparse identification of nonlinear dynamics (SINDy) algorithm~\cite{Brunton16a,Brunton16}, which infers governing equations directly from data. As a first step toward this broader objective, we focus here on a more tractable problem.

In this paper, we consider networks for which the exact mean-field model in the thermodynamic limit is known, but its parameters are unknown. To emulate realistic experimental conditions, we aim to infer these parameters from data obtained through partial observations of a finite-size network. Specifically, we assume that only a single scalar macroscopic variable is accessible, while the remaining macroscopic variables entering the mean-field model are unobserved (hidden). Our objective is twofold: (i) to estimate the parameters of the mean-field model, and (ii) to reconstruct the dynamics of the hidden macroscopic variables from limited observations.

There is no universal inference algorithm suitable for all nonlinear dynamical systems; consequently, a wide range of parameter estimation methods has been developed~\cite{Chou2009,Sun2012,Petrowski2017,Cranmer2020,Roda20,Niven24,dupont19,Kong_2022,Somacal22}. An important class consists of derivative-free global optimization techniques, including simulated annealing~\cite{Kirkpatrick1984,Dai2014}, bio-inspired metaheuristics such as ant-stigmergy~\cite{Korosec2012,Tashkova2012} and particle swarm optimization~\cite{Bonyadi2017,Akman2018,Baldy24}, as well as evolutionary algorithms including genetic algorithms~\cite{Holland1994,Sun2012}, differential evolution (DE)~\cite{Storn1997,Sun2012,Baldy24}, and the covariance matrix adaptation evolution strategy~\cite{Hansen1996}. Comparative studies indicate that DE is particularly efficient for parameter inference in nonlinear systems~\cite{Tashkova2012,Baldy24}. DE employs vector-based mutation and recombination strategies to generate candidate solutions, enabling effective exploration of high-dimensional parameter spaces. In addition to its strong performance, DE is straightforward to implement, requires relatively few control parameters, and exhibits robustness to moderate noise. Its population-based structure also facilitates parallel computation, which is advantageous for computationally intensive inference tasks.

In this paper, we therefore employ DE to infer parameters in two neural network models, referred to as QIF-IN and QIF-AD networks. The QIF-IN model, introduced by Devalle et al.~\cite{Devalle17}, describes a heterogeneous population of inhibitory QIF neurons with synaptic kinetics and exhibits periodic collective oscillations. The QIF-AD model, proposed by Pietras et al.~\cite{montbrio2025}, represents a heterogeneous population of excitatory QIF neurons with spike-frequency adaptation and displays chaotic collective dynamics. Both models admit exact reductions to low-dimensional mean-field equations in the thermodynamic limit.

The remainder of the paper is organized as follows. In Sec.~\ref{sec:problem}, we formulate the inference problem and define the loss function used to estimate the mean-field parameters. Section~\ref{sec:methods} introduces two synchronization-based approaches that remove the dependence of the loss function on the unknown initial conditions of the hidden macroscopic variables, and describes the DE algorithm. Section~\ref{sec:results} presents the results of parameter estimation and hidden-variable reconstruction for the QIF-IN and QIF-AD networks. These findings are discussed in Sec.~\ref{sec:disc}. Finally, Sec.~\ref{sec:models} provides detailed descriptions of the microscopic and mean-field formulations of the QIF-IN and QIF-AD models.

\section{Problem statement}\label{sec:problem}

We consider a heterogeneous network of $N$ coupled neurons whose microscopic dynamics are governed by a system of ordinary differential equations:
\begin{equation}
\dot{\bold{x}}_j(t) = \bold{f} \left[\bold{x}_1(t),\ldots, \bold{x}_N(t),  \bold{p}_j, I_\mathrm{ext}(t) \right], \quad j=1,\ldots, N.
\label{eq:micro_model}
\end{equation}
Here $\bold{x}_j(t)$ denotes the state vector of the $j$-th neuron, $\bold{p}_j$ is a heterogeneous parameter vector specifying its intrinsic properties, and $I_{\mathrm{ext}}(t)$ is a homogeneous external input current. In experimental settings, microscopic variables are typically inaccessible, and only macroscopic observables describing collective network activity, such as the mean membrane potential or mean firing rate, can be measured. We assume that the available data consist of a time series of a single scalar macroscopic variable, which we denote as $X_\mathrm{out}(t)$.

Direct simulation of the microscopic system~\eqref{eq:micro_model} in order to reproduce the observed signal $X_\mathrm{out}(t)$ is computationally demanding when $N$ is large and individual neurons are described by complex equations. Even if the analytical form of the vector field $\bold{f}$ is known, fitting the heterogeneous parameters $\bold{p}_j$ from a single scalar observable is practically infeasible due to the extremely high dimensionality of the parameter space. An alternative approach is to employ low-dimensional neural mass models~\cite{Wilson1972,Destexhe2009}. Classical neural mass models significantly reduce computational complexity but are phenomenological and generally capture only qualitative features of macroscopic dynamics. Recent advances in nonlinear dynamics have led to the development of next-generation neural population models~\cite{Coombes2023}, which are derived directly from the microscopic equations under the assumption that trajectories collapse onto a low-dimensional invariant manifold. This approach was first demonstrated by Ott and Antonsen~\cite{Ott2008} for heterogeneous populations of Kuramoto oscillators and later extended to neuroscience. In particular, Montbri\'{o} et al.~\cite{Montbrio2015} obtained an exact low-dimensional mean-field description for infinite networks of pulse-coupled quadratic integrate-and-fire neurons. Since then, NGNPM research has become a hot topic in theoretical neuroscience~\cite{Coombes2023}.

In the present work, we assume that the microscopic model~\eqref{eq:micro_model} admits such a low-dimensional invariant manifold and can therefore be reduced, in the thermodynamic limit $N \to \infty$, to a closed mean-field system
\begin{equation}
\dot{\bold{X}}(t) = \bold{F} \left[\bold{X}(t),\bold{P},I_\mathrm{ext}(t) \right],
\label{eq:model_macr}
\end{equation}
where $\bold{X}(t)$ is a low-dimensional vector of macroscopic variables and $\bold{P}$ is a vector of mean-field parameters governing the collective dynamics. We assume that one component $X_m(t)$ of $\bold{X}(t)$ corresponds to the output $X_\mathrm{out}(t)$ of the of the microscopic model~\eqref{eq:micro_model}, which we interpret here as an experimentally observed signal of a real network.

Our objective is to estimate the parameter vector $\bold{P}$ from partial observations of a finite-size network. Specifically, we determine $\bold{P}$ by requiring that the model output $X_m(t)$ reproduces the measured signal $X_\mathrm{out}(t)$ as accurately as possible. To mimic experimental conditions, we consider discrete measurements over a finite time interval, $X_\mathrm{out}(t_k)$, $k=1,\ldots,M$, with uniform sampling step $\delta t = t_{k+1}-t_k$. We define the loss function
\begin{equation}
L(\bold{P},\bold{X}(0)) = \frac{1}{2M}\sum_{k=1}^M 
\left\{ X_m[t_k;\bold{P},I_\mathrm{ext}(t_k),\bold{X}(0)] 
- X_\mathrm{out}(t_k) \right\}^2,
\label{eq:loss_function}
\end{equation}
where the notation emphasizes that $X_m(t_k)$ is obtained by integrating the mean-field system~\eqref{eq:model_macr} with parameters $\bold{P}$ and initial condition $\bold{X}(0)$. Minimization of $L$ yields the parameter vector $\bold{P}$ that ensures optimal consistency between the mean-field model and the observed scalar time series. Importantly, the loss function depends on the full initial state $\bold{X}(0)$, whereas experimentally only the initial value $X_m(0)$ of the observed component is available. The initial conditions of the unobserved (hidden) macroscopic variables must therefore be estimated jointly with the parameters $\bold{P}$. This joint estimation problem becomes particularly challenging in the presence of chaotic dynamics, where sensitivity to initial conditions strongly affects trajectory matching. An additional complication arises from finite-size effects. The mean-field system~\eqref{eq:model_macr} is exact only in the limit $N \to \infty$, whereas the measured signal $X_\mathrm{out}(t)$ originates from a finite network. Formally, $X_\mathrm{out}(t)$ can be viewed as the sum of the deterministic mean-field trajectory $X_m(t)$ and fluctuations induced by finite-size noise~\cite{Klinshov2022,Kirillov2024}. Consequently, even for optimal parameters and initial conditions, perfect trajectory matching is generally impossible. We therefore analyze how finite-size fluctuations influence parameter inference accuracy as a function of network size $N$.

To summarize,  the central objective of this work is to develop suitable  algorithms for the minimization of the loss function~\eqref{eq:loss_function} and to infer the parameters of the mean-field model~\eqref{eq:model_macr} and time series of the hidden macroscopic variables  using the time series of only one scalar macroscopic variable measured in a finite-size network~\eqref{eq:micro_model}.  We demonstrate the  performance of our algorithms using  the QIF-IN~\cite{Devalle17} and QIF-AD~\cite{montbrio2025} models that exhibit periodic and chaotic collective oscillations, respectively. Both models admit exact reduction to low-dimensional mean-field equations in the thermodynamic limit. A detailed description of these models is provided in Sec.~\ref{sec:models}.

\section{Methods}\label{sec:methods}

When minimizing the loss function~\eqref{eq:loss_function}, the main difficulty is associated with its dependence on the initial conditions of the unobservable components of the state vector $\bold{X}(t)$. This problem becomes particularly challenging when the network is in chaotic mode and its trajectories are extremely sensitive to small changes in initial conditions. To overcome this issue, we formulate the optimization procedure by exploiting the phenomenon of synchronization. Specifically, by employing appropriate coupling strategies, we induce synchronization between the trajectories of the mean-field model and the output $X_\mathrm{out}(t)$ of the microscopic model, thereby eliminating the need to fit unknown initial conditions. In the synchronized regime, the mean-field system effectively ``forgets'' its initial state and subsequently follows the dynamics of the output variable $X_\mathrm{out}(t)$. As a result, the asymptotic dynamics of the mean-field system become independent of its initial conditions.

In practice, to ensure that the loss function is not influenced by transient effects associated with initialization, we evaluate it only after a transient time $t_\mathrm{trans}$ that exceeds the characteristic synchronization time. Using this approach, we solve the optimization problem via two distinct synchronization-based methods, which we refer to as noninvasive and invasive.

\subsection{Noninvasive method}

The first synchronization method we use is noninvasive in the sense that it does not require any external intervention in the observed system. In particular, we do not affect the neural network with external current, setting 
\begin{equation}
I_\mathrm{ext}(t)=0.
\label{eq:Iext0}
\end{equation}
The dynamics of the mean-field model  is guided toward the measured trajectory using a master-slave control scheme~\cite{Parlitz96,Yang06,quinn09,abarbanel09}, where the master corresponds to the signal $X_\mathrm{out}(t)$  from the actual network and the slave represents the inferred mean-field model.
Specifically, we modify the mean-field system~\eqref{eq:model_macr} by adding a coupling term
\begin{equation}
\dot{\bold{X}}(t) = \bold{F} \left[\bold{X}(t),\bold{P},0 \right]+K\bold{I} \, [X_\mathrm{out}(t)-X_m(t)],
\label{eq:model_macr_feed}
\end{equation}
where $K$ denotes the feedback gain coefficient, and $\bold{I}$ is a unit vector of the same length as $\bold{X}$, all components of which are zero except for the component $m$. In a more general formulation, the vector $\bold{I}$ can be replaced by an arbitrary coupling matrix of the corresponding dimension, but here we restrict ourselves to the simplest case, when the coupling term applies only to the $m$-th equation defining the dynamics of the $X_m$-component. For $K>0$, this term provides negative feedback that drives $X_m(t)$ towards the observed trajectory $X_\mathrm{out}(t)$, thereby reducing the mismatch. We assume that for the corresponding values of $K$, generalized synchronization~\cite{Rulkov1995,Pyragas1998} occurs, i.e., all conditional (transverse) Lyapunov exponents (LEs) of the system~\eqref{eq:model_macr_feed} become negative. Since the parameter vector $\bold{P}$ changes during the optimization process, we also assume that the synchronization condition remains satisfied in the vicinity of the optimal value $\bold{P}$. 

The analysis of the synchronization phenomenon for the specific models considered in this paper is carried out in Sec~\ref{sec:models}.  Figures~\ref{fig:arnold_lyap_dev} (a) and \ref{fig:arnold_lyap_mont} (a) illustrate the dependence of the maximum conditional LE on the feedback coefficient $K$ for the QIF-IN and QIF-AD networks, respectively.
Without feedback ($K=0$), the QIF-IN network exhibits periodic oscillations and its maximum LE value is zero. Even a small negative feedback (K>0) makes this exponent negative and causes generalized synchronization. The QIF-AD network is initially chaotic and its maximum LE is positive. To make this exponent negative, higher values of the feedback coefficient $K$ are required.

Note that, in the numerical integration of the mean-field Eqs.~\eqref{eq:model_macr_feed}, the variable $X_\mathrm{out}(t)$ is available only at discrete time instants $t=t_k$. We assume that the sampling interval $\delta t$ is sufficiently small so that interpolation of $X_\mathrm{out}(t)$ yields a stable solution. 

\subsection{Invasive method}

The second method we propose is invasive, since in this case the observed system is subject to perturbation. Here we achieve synchronization of the mean-field model~\eqref{eq:model_macr} with the observed network~\eqref{eq:micro_model}  by applying a periodic external current $I_{\mathrm{ext}}(t)$. We choose the waveform of this current, similar to that used in Ref.~\cite{Devalle17}:
\begin{equation}
I_{\mathrm{ext}}(t) = K[1+\sin(2\pi t/T_\mathrm{ext})/2]^3,\label{eq:periodicF}
\end{equation}
which describes excitatory (K>0) or inhibitory (K<0) periodic pulses. The parameter $K$ determines the pulse amplitude, and $T_\mathrm{ext}$ determines the period. By selecting appropriate values for these parameters, both the actual network~\eqref{eq:micro_model} and the mean-field model~\eqref{eq:model_macr} can be synchronized with the external current. Examples of synchronization regions in the $(T_\mathrm{ext},K)$ parameter plane, known as Arnold tongues, are shown in Figs.~\ref{fig:arnold_lyap_dev} (b) and \ref{fig:arnold_lyap_mont} (b) for the QIF-IN and QIF-AD networks, respectively. For both networks, inhibitory pulses provide a larger area of Arnold tongues, so we chose inhibitory pulses in our parameter estimation algorithm to make the optimization process more robust. It should be noted that in the case of an initially chaotic mode, a periodic external current transfers the network to a periodic mode and makes the system less sensitive to the initial conditions. Due to this property, the invasive method may be preferable to the noninvasive one. 

\subsection{Optimization algorithm}

In preliminary numerical experiments, we compared several optimization approaches, including gradient-based methods, particle swarm optimization, and the covariance matrix adaptation evolution strategy. Among them, the DE algorithm consistently demonstrated the most accurate parameter recovery and the fastest convergence, so we chose it as the main tool in our research.

DE is a stochastic, population-based optimization algorithm well suited for continuous, nonlinear, and non-differentiable problems. It maintains a population of candidate solutions that evolves over successive generations, with predefined lower and upper search bounds for all optimized parameters. The initial population is randomly generated to cover the entire parameter space within these bounds.  The algorithm proceeds through three stages: mutation, crossover, and selection. During the mutation stage, a new parameter vector, referred to as the mutant vector, is generated by adding a weighted difference between two population vectors to a third vector. In the crossover stage, a target vector is selected, and components of the mutant vector are randomly exchanged with the corresponding components of the target vector to produce a trial vector. In the selection stage, the loss function is evaluated for both the trial and target vectors, and the vector with the smaller loss is retained for the next generation.

In our study, we employ the ``best1bin'' strategy~\cite{SciPy20}, in which mutation is guided by the current best solution in the population, promoting efficient convergence. The population size is set to $15 \times N_p$, where $N_p$ denotes the number of parameters to be estimated. The best solution is updated once per generation, following the standard DE selection procedure. All simulations use the Python implementation of DE provided by SciPy~\cite{SciPy20}.

\section{Results\label{sec:results}}

We apply the DE optimization algorithm to scalar time series generated by two different types of neural networks, QIF-IN and QIF-AD, described in detail in Secs.~\ref{sec:period} and \ref{sec:chaos}, respectively.
We compare DE optimization capability to infer the parameters of the corresponding mean-field equations using both noninvasive and invasive  synchronization strategies for different network sizes $N=2\cdot 10^2,5\cdot 10^2,10^3,10^4,10^5$, and $N\to \infty$. The infinite-size network time series is generated by the mean-field model, which is exact in the thermodynamic limit $N\to \infty$.

\subsection{QIF-IN network\label{sec:QIF-IN_network}}

\begin{figure}
\includegraphics[width=0.95\textwidth]{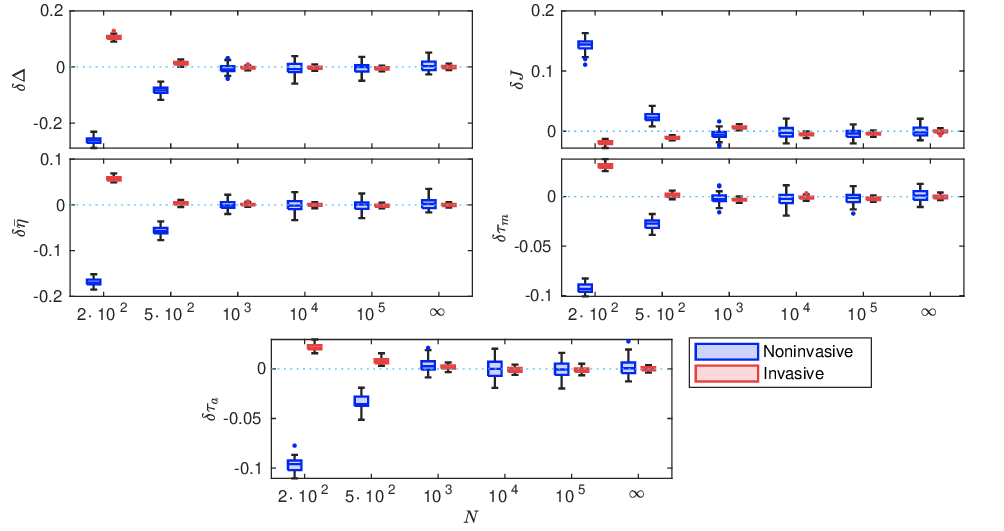}
\centering
\caption{\label{fig:QIF_IN_param} Inference of mean-field model parameters from chaotic time series of mean membrane potential observed in QIF-IN networks of different sizes. Box plots visualize the statistics of relative deviations of inferred parameters from their actual values used in the QIF-IN microscopic network model. The statistics are based on $40$ different initial sets of mean-field model parameters, randomly selected from the optimization bounds presented in Tab.~\ref{tab_Devalle}. Blue and red colors represent the noninvasive and invasive synchronization methods, respectively. 
For the noninvasive method, the transient time is $t_\mathrm{trans}=831.3$~ms, the loss function is optimized over a training interval of $t_\mathrm{train}=277.1$~ms and the feedback strength is $K=0.5$. For the invasive method, the parameters are: $t_\mathrm{trans}=1400$~ms, $t_\mathrm{train}=560$~ms, $K=-0.45$ and $T_\mathrm{ext}=28$~ms. The remaining parameters are given in the Tab.~\ref{tab_Devalle}.
}
\end{figure}

\begin{figure}
\includegraphics{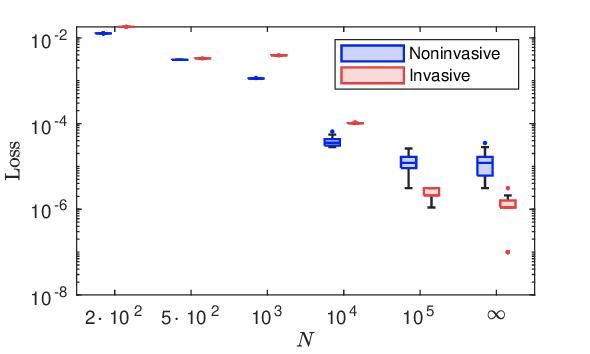}
\centering
\caption{\label{fig:QIF_IN_Loss} Box plot of the loss function minima for different sizes of the QIF-IN network. Noninvasive and invasive methods are represented in blue and red colors, respectively. The parameter values are given in the caption of Fig.~\ref{fig:QIF_IN_param} and in Tab.~\ref{tab_Devalle}.
}
\end{figure}

\begin{figure}
\includegraphics{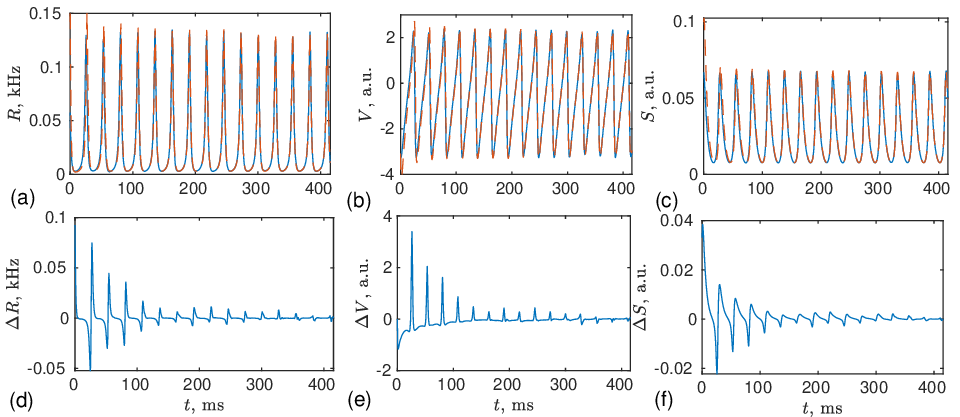}
\centering
\caption{\label{fig:QIF_IN_pred} Reconstruction of the dynamics of unobservable macroscopic variables $R$ and $S$ of  QIF-IN network consisting of 1000 neurons. (a), (b) and (c) The dynamics of the mean firing rate $R$, mean membrane potential $V$ and synaptic activation variable $S$, respectively. 
The blue curves correspond to the microscopic model, while red dashed curves show the dynamics of the mean-field model, which is driven by the mean membrane potential of the microscopic model via a master-salve control scheme. The parameters of the mean-field model are set to the median values obtained from  optimization using the DE method with noninvasive synchronization. (d), (e) and (f) Error dynamics showing the differences between macroscopic variables estimated from a 1000-neuron network and the corresponding mean-field model variables.
}
\end{figure}

The results for the QIF-IN network that exhibits periodic collective oscillations are presented in Figs.~\ref{fig:QIF_IN_param} -- \ref{fig:QIF_IN_pred}. On the microscopic level, the network dynamics is described by $N+1$ differential Eqs.~\eqref{eq:micrDevalle}. The dynamics of individual neurons are defined by their membrane potentials $v_j(t)$, $j=1,\ldots, N$. The neurons are inhibitory coupled trough a global variable $S(t)$ that  describes the dynamics of synaptic activation. The mean membrane potential of the microscopic model is interpreted as the experimentally observed output signal $X_\mathrm{out}(t)$ of the network. 
In the thermodynamic limit $N\to \infty$, the microscopic model reduces to a closed system of mean-field Eqs.~\eqref{eq:macrDevalle} that describe the dynamics of  three macroscopic variables: the mean firing rate $R(t)$, the mean membrane  potential $V(t)$, and the synaptic activation variable $S(t)$. For this network, the variable $X_m(t)$ in Eqs.~\eqref{eq:loss_function}, \eqref{eq:model_macr} and \eqref{eq:model_macr_feed} is defined as $X_m(t):=X_2(t):=V(t)$. The mean-field model has five independent parameters:   $\Delta$, $\bar{\eta}$, $J$, $\tau_m$ and $\tau_d$. The physical meaning of these parameters is described in Sec.~\ref{sec:period}, and their actual values used to generate the output $X_\mathrm{out}(t)$ of the microscopic model Eq.~\eqref{eq:micrDevalle} are presented in Tab.~\ref{tab_Devalle}.


Figure~\ref{fig:QIF_IN_param} shows the relative errors for all five inferred parameters, i.e, their relative deviations from the actual values used in the microscopic network model. Results are presented for six different values of the network size $N$. For each $N$, $40$ different randomly chosen initial values of the parameters were taken from the corresponding optimization bounds presented in Tab.~\ref{tab_Devalle}.
Summary statistics of relative deviations are visualized using box plots. Blue and red colors represent the noninvasive and invasive synchronization methods, respectively. The errors for both methods are spread over a small interval, which means that the DE algorithm converges to almost the same minimum of the loss function regardless of the initial choice of parameter values. The invasive method gives a slightly smaller spread of errors than the noninvasive method. In both cases, the relative errors for all parameters decrease with increasing network size, and for $N \geq 1000$, both methods provide estimates of the actual parameter values with a relative accuracy better than $1\%$. Figure \ref{fig:QIF_IN_Loss} complements these results with a graph of the loss function minimum. As the network size increases, the minimum of the loss function for both methods decreases due to the reduced influence of finite-size effects (finite-size noise). 

Finally, Fig.~\ref{fig:QIF_IN_pred} demonstrates that the mean-field model with parameters obtained from the observation of only one macroscopic variable can successfully reconstruct the dynamics of the other macroscopic variables, which, according to our assumption, are not accessible to experimental observation. 
We set the parameters of the mean-field model to the median values obtained by noninvasive method using time series of the mean membrane potential of a microscopic model consisting of $N=1000$ neurons. Figures~\ref{fig:QIF_IN_pred}(a), \ref{fig:QIF_IN_pred}(b) and \ref{fig:QIF_IN_pred}(c) show the dynamics of the mean firing rate $R(t)$, the mean membrane potential $V(t)$ and the synaptic activation variable $S(t)$, respectively. 
The blue curves correspond to a microscopic model of size $N=1000$, and the red dashed curves are obtained from a mean-field model driven by the mean membrane potential of the microscopic model. Figures ~\ref{fig:QIF_IN_pred}(d), \ref{fig:QIF_IN_pred}(e), and \ref{fig:QIF_IN_pred}(f) show the dynamics of the differences between the macroscopic variables calculated from the microscopic model and the corresponding variables predicted by the driven mean-field model. The initial conditions for the mean-field equations were chosen randomly, with the exception of the mean membrane potential, the initial value of which was obtained from the microscopic model. After the transient process, the mean-field model synchronizes with the microscopic model and not only reproduces the observed variable $V(t)$, but also correctly reconstructs the dynamics of the unobserved macroscopic variables $R(t)$ and $S(t)$ of the microscopic model. In the post-transient period, errors in the mean firing rate $\Delta R$, membrane potential $\Delta V$ and  activation variable $\Delta S$ fluctuate around zero due to finite-size noise.

\subsection{QIF-AD network\label{sec:QIF-AD_network}}

\begin{figure}
\includegraphics[width=0.95\textwidth]{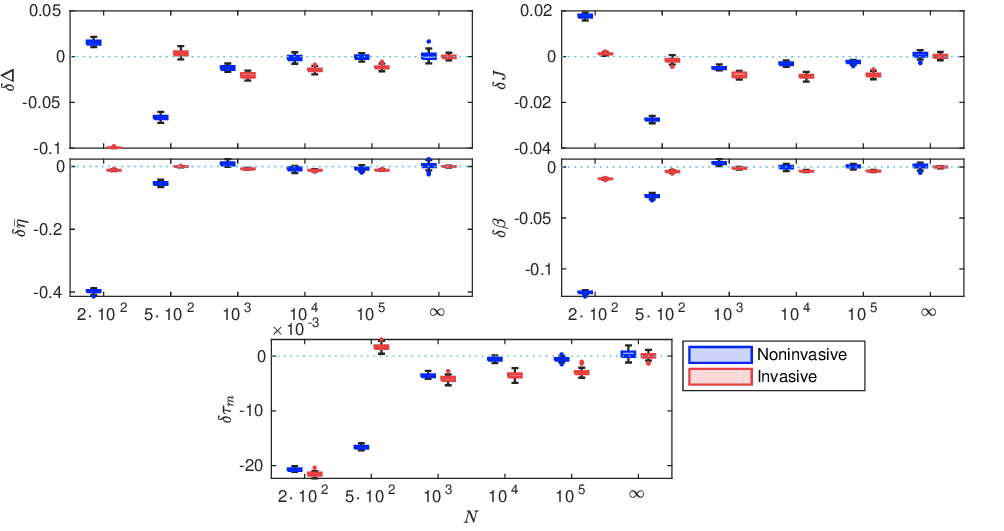}
\centering
\caption{\label{fig:QIF_adapt_param}
Inference of mean-field model parameters from chaotic time series of mean membrane potential observed in QIF-AD networks of different sizes. Box plots visualize the statistics of relative deviations of inferred parameters from their actual values used in the QIF-AD microscopic network model. The statistics are based on $40$ different initial sets of mean-field model parameters, randomly selected from the optimization bounds presented in Tab.~\ref{tab_Mont}. For the noninvasive method, the transient time is $t_\mathrm{trans}=1000$~ms, the loss function is optimized over a training interval of $t_\mathrm{train}=500$~ms and the feedback strength is $K=5$. For the invasive method, the parameters are: $t_\mathrm{trans}=2400$~ms, $t_\mathrm{train}=240$~ms, $K=-4$ and $T_\mathrm{ext}=80$~ms. The remaining parameters are given in the Tab.~\ref{tab_Mont}.
}
\end{figure}

\begin{figure}
\includegraphics{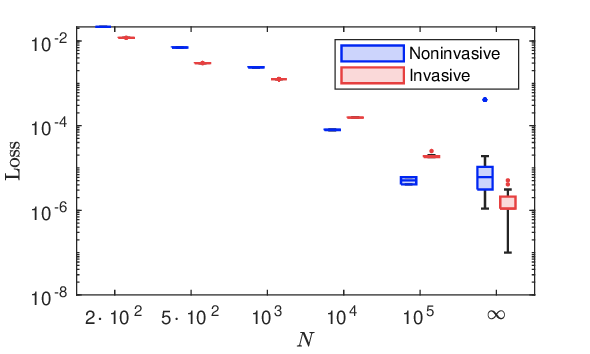}
\centering
\caption{\label{fig:QIF_adapt_Loss} 
Box plot of the loss function minima for different sizes of the QIF-AD network. The parameter values are given in the caption of Fig.~\ref{fig:QIF_adapt_param} and in Tab.~\ref{tab_Mont}.
}
\end{figure}

\begin{figure}
\includegraphics{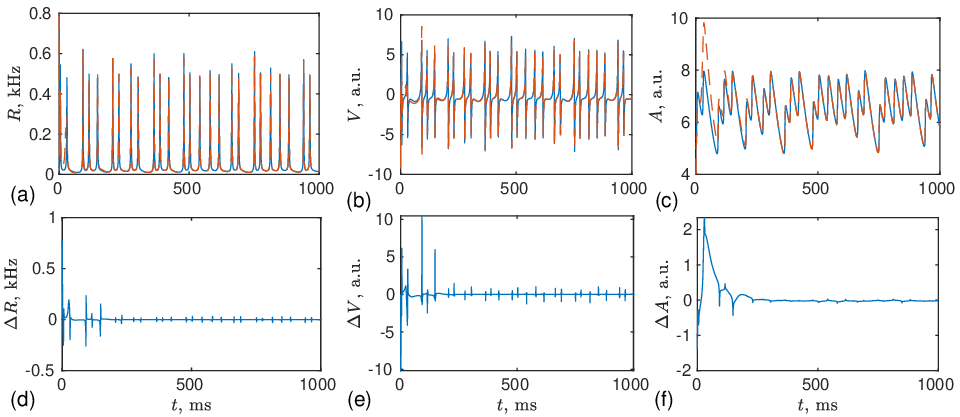}
\centering
\caption{\label{fig:QIF_adapt_pred} 
Reconstruction of the chaotic dynamics of unobservable macroscopic variables $R$ and $A$ of  QIF-AD network consisting of 1000 neurons. (a), (b) and (c) The dynamics of the mean firing rate $R$, mean membrane potential $V$ and mean adaptation variable $A$, respectively. 
The blue curves correspond to the microscopic model, while red dashed curves show the dynamics of the mean-field model, which is driven by the mean membrane potential of the microscopic model via a master-salve control scheme. The parameters of the mean-field model are set to the median values obtained from  optimization using the DE method with noninvasive synchronization. (d), (e) and (f) Error dynamics showing the differences between macroscopic variables estimated from a 1000-neuron network and the corresponding mean-field model variables.
}
\end{figure}

The problem of determining the parameters of the mean field model for the QIF-AD network is more complex, since here we are considering a regime of chaotic collective oscillations. On the microscopic level, the dynamics of the QIF-AD network  consisting of $N$ neurons is described by $2N$ differential Eqs.~\eqref{eq:micrMontbrio}. The dynamics of individual neurons are defined by their membrane potentials $v_j(t)$ and adaptation variables $a_j(t)$, $j=1,\ldots, N$. The neurons interact with each other via an excitatory mean-field coupling. As in the previous case, the mean membrane potential of the microscopic model is interpreted as the experimentally observed output signal $X_\mathrm{out}(t)$ of the network. In the thermodynamic limit $N\to \infty$, this model reduces to a three-dimensional system of mean-field Eqs.~\eqref{eq:macrMontbrio} defined by three macroscopic variables: the mean firing rate $R(t)$, the mean membrane  potential $V(t)$, and the mean adaptation variable $A(t)$. The variable $X_m(t)$ in Eqs.~\eqref{eq:loss_function}, \eqref{eq:model_macr} and \eqref{eq:model_macr_feed} is defined as $X_m(t):=X_2(t):=V(t)$. This model has six independent parameters:   $\Delta$, $\bar{\eta}$, $J$, $\beta$, $\tau_m$ and $\tau_a$ (see Sec.~\ref{sec:chaos} for their physical meaning). The actual values of these parameters used in the microscopic model Eq.~\eqref{eq:micrMontbrio}, as well as their optimization boundaries used in the DE algorithm, are presented in Tab.~\ref{tab_Mont}. To reduce the computation time of the DE optimization algorithm, we fix the value of the parameter $\tau_a$ and consider the remaining five parameters as unknown parameters whose values should be determined through optimization of the loss function. The results of numerical simulation for the QIF-AD network are presented in Figs.~\ref{fig:QIF_adapt_param} -- \ref{fig:QIF_adapt_pred}.

Figure~\ref{fig:QIF_adapt_param} shows the statistics of relative errors for all inferred parameters depending on the network size $N$.  
The statistics are based on $40$ experiments with different initial parameter values randomly selected from the corresponding optimization bounds presented in Table ~\ref{tab_Mont}. As in the previous case, relative errors decrease with increasing network size for both noninvasive and invasive methods. At $N \geq 1000$, the finite-size noise becomes small enough that both methods can provide estimates of the actual values of the mean-field model parameters with a relative accuracy better than $1\%$. Figure~\ref{fig:QIF_adapt_Loss} shows the decrease in the minimum of the loss function with increasing network size for both methods. Figure~\ref{fig:QIF_adapt_pred} demonstrates the ability of our approach to recover the dynamics of unobserved macroscopic network variables even when the network exhibits chaotic oscillations. The mean-field model with median values of the parameters estimated during the optimization procedure is controlled by the mean membrane potential of the microscopic model of size $N=1000$ via a master-slave coupling. The initial condition for the mean membrane potential $V$ of the mean-field model were taken from the observed time series of the microscopic model, while the initial conditions for the unobserved  macroscopic variables $R$ and $A$ were chosen randomly. After the transient process, the mean-field model synchronizes with the microscopic model and correctly reconstructs the chaotic dynamics of the unobservable macroscopic variables $R(t)$ and $A(t)$ [see Figs.~\ref{fig:QIF_adapt_pred}(a),\ref{fig:QIF_adapt_pred}(d)) and \ref{fig:QIF_adapt_pred}(c),\ref{fig:QIF_adapt_pred}(f), respectively].

\section{Discussion}\label{sec:disc}
 
In this study, we examined how well the parameters  of mean-field equations, which are valid in the thermodynamic limit, can be inferred from partial observations of a finite-size neural network. Among various available optimization strategies, we employed the differential evolution (DE) algorithm~\cite{Storn1997,Sun2012,Baldy24}. As a derivative-free global optimization method with a simple implementation and few control parameters, DE demonstrated robust and efficient performance in the inference of model parameters and hidden macroscopic variables.

To reduce the dimensionality and complexity of the optimization problem, we removed the explicit dependence of the loss function on the unknown initial conditions of the mean-field variables. This was accomplished using two synchronization-based approaches, referred to as noninvasive and invasive methods. The noninvasive method leaves the original neural network dynamics unchanged and relies on a master-slave coupling scheme to synchronize the mean-field model with the observed network output. In contrast, the invasive method introduces a common external periodic perturbation to both the real network and the mean-field model to achieve synchronization. Such synchronization-based strategies are particularly important for chaotic dynamics, where the sensitivity of the mean-field model becomes extremely high to the choice of initial conditions.

Application of the proposed framework to the periodically oscillating QIF-IN network and the chaotically oscillating QIF-AD network yielded qualitatively similar results. In both cases, we successfully inferred the parameters of the mean-field model using a time series of a single macroscopic observable -- the mean membrane potential of a finite-size network. The relative parameter estimation error decreases with increasing network size and falls below $1\%$ for networks larger than approximately 1000 neurons. This indicates that the DE-based inference procedure remains reliable even for moderately sized networks, where finite-size fluctuations are still significant.

In addition to parameter estimation, the dynamics of hidden macroscopic variables were accurately reconstructed for both periodic and chaotic regimes in networks with as few as 1000 neurons. Notably, the DE algorithm exhibited a small spread of errors across different initial guesses of the fitting parameters, suggesting convergence toward a consistent minimum of the loss function. The invasive synchronization method provided a slightly narrower error distribution compared to the noninvasive approach, indicating improved robustness in parameter recovery.

The demonstrated ability to infer mean-field parameters and reconstruct hidden macroscopic dynamics from limited observations of finite-size networks establishes a foundation for more general data-driven approaches in neural population modeling. A natural next step is to move beyond parameter inference within a known mean-field structure and address the more challenging problem of reconstructing the governing mean-field equations themselves. Such developments could extend NGNPM theory~\cite{Coombes2023} beyond the class of QIF neurons.

One promising direction is the sparse identification of nonlinear dynamics framework~\cite{Brunton16a,Brunton16}, which aims to learn governing differential equations directly from data. Standard SINDy implementations typically assume full observability of system variables. However, recent work by Bakarji et al.~\cite{Bakarji2023} demonstrated that combining deep learning with SINDy enables the identification of interpretable low-dimensional models in effective coordinates, even when only a single scalar variable of a chaotic system is observable. Their results, illustrated for the R\"ossler and Lorenz systems, suggest that similar strategies may be applicable to neural population models. Extending such data-driven equation discovery techniques to heterogeneous neural networks composed of different neuron types constitutes an important direction for future research.

\section{Models}\label{sec:models}

We consider two neural network models labeled as QIF-IN and QIF-AD networks. Both are composed of QIF neurons and admit reduction to exact low-dimensional mean-field equations in the thermodynamic limit. The QIF-IN network exhibits periodic collective oscillations, while the QIF-AD network demonstrates chaotic collective dynamics. Below we briefly present mathematical description of these models on the microscopic level and their reduced mean-field representations.

\subsection{Microscopic and mean-field models of the QIF-IN network}\label{sec:period}

Microscopic QIF-IN model describes a heterogeneous population of $j=1\ldots N$ QIF neurons interacting via a mean-field inhibitory coupling. The evolution of the membrane potential $v_j$ of a neuron $j$ in the ensemble and the synaptic activation $S$ defining the coupling is described by the differential equations~\cite{Devalle17}: 
\begin{subequations}
\label{eq:micrDevalle}
\begin{eqnarray}
\tau_m \dot{v}_j & = & v^2_j + \eta_j - J \tau_m S+ I_\mathrm{ext}(t),
 \label{eq:micrDevalleV}\\
\tau_d \dot{S} & = &  -S + R \label{eq:micrDevalleS}
\end{eqnarray}
\end{subequations}
with the auxiliary resetting rule: each time a potential $v_j$ reaches the infinity, it is reset to minus infinity, and the neuron emits an instantaneous spike which contributes to the network mean firing rate: 
\begin{equation}
R(t) = \lim_{\tau \to 0} \frac{1}{N} \sum_{j=1} ^N \sum_k \frac{1}{\tau} \int_{t-\tau} ^{t} \delta(s-t^k_j) \mathrm{d}s. 
\label{eq:R(t)}
\end{equation}
Here $t^k_j$ is the time of the $k$th spike of $j$th neuron, and $\delta(t)$ is the Dirac delta function. The parameters $\tau_m$ and $\tau_d$ are the membrane and synaptic time constants, respectively. The heterogeneity parameter $\eta_j$ is a constant current that determines the behavior of each isolated neuron. The values of these parameters are generated according to a Lorentzian density function centered at $\bar{\eta}$ and having half-width $\Delta$:
\begin{equation}
g(\eta) = \frac{1}{\pi} \frac{\Delta}{(\eta-\bar{\eta})^2+\Delta^2}.
 \label{eq:Lorentz}
\end{equation}
The positive parameter $J$ determines the strength of the inhibitory coupling, and $I_\mathrm{ext}(t)$ represents the external uniform current.

Devalle et al.~\cite{Devalle17} showed that in the thermodynamic limit $N \to \infty$ the microscopic model~\eqref{eq:micrDevalle} can be reduced to an exact closed system of three differential equations
\begin{subequations}
\label{eq:macrDevalle}
\begin{eqnarray}
\tau_m \dot{R} & =& \frac{\Delta}{\pi \tau_m}+2RV, \label{eq:macrDevalleR}\\
\tau_m \dot{V} & =& V^2-(\tau_m \pi R)^2 + \bar{\eta} - J \tau_m S + I_\mathrm{ext}(t), \label{eq:macrDevalleV}\\
\tau_d \dot{S} & =&  -S + R, \label{eq:macrDevalleS}
\end{eqnarray}
\end{subequations}
which defines the dynamics of three macroscopic variables: the mean firing rate $R$, the mean membrane potential $V$, and the mean synaptic activation $S$. This model exhibits periodic collective oscillations over a wide range of parameter values. The specific parameter values (actual parameters) used in our simulations are listed in Tab.~\ref{tab_Devalle}.
\begin{table}[h]
\caption{\label{tab_Devalle} \textbf{Values of QIF-IN network parameters and optimization bounds used in the DE algorithm.} 
}
\centering
\rowcolors{2}{gray!10}{white} 
\begin{tabular}{cccc}
\rowcolor{gray!30}Parameters 	& Actual values 	&  Optimization bounds		& Units \\
$\Delta$ 	& 0.3				& $[0.07,\; 0.7]$ 	& a.u.	\\
$\bar{\eta}$& 4					& $[1.75,\; 4.9]$	& a.u.	\\
$J$ 			& 21 				& $[10,\; 30]$		& a.u.	\\
$\tau_m$ 	& 10					& $[0.25,\; 15]$		& ms	\\
$\tau_d$ 	& 5 					& $[1,\; 17]$		& ms
\end{tabular}
\end{table}

\begin{figure}
\includegraphics[width=0.95\textwidth]{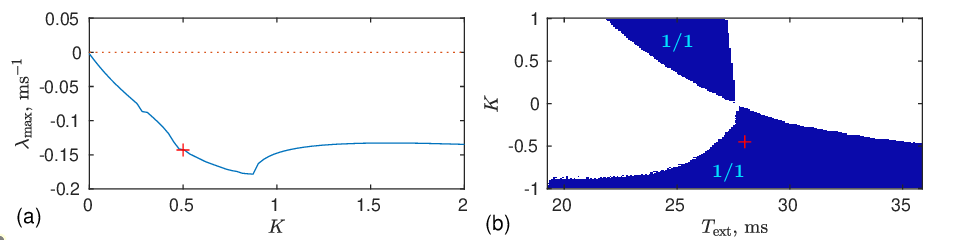}
\centering
\caption{\label{fig:arnold_lyap_dev} Synchronization regions for the QIF-IN network. (a) Maximal conditional Lyapunov exponent as a function of the feedback gain $K$ of the noninvasive method. (b) Arnold tongues in the parameter plane $(T_\mathrm{ext}, K)$ for the invasive method.  The blue color indicates the synchronization regions of the mean-field model  Eq.~\eqref{eq:macrDevalle} with an external periodic current Eq.~\eqref{eq:periodicF} of amplitude $K$ and period $T_\mathrm{ext}$. Arnold tongues are labeled with winding numbers $w=1/1$, which means that the network generates one spike per period of the external current. The red crosses in panels (a) $K =0.5$ and (b) $(T_\mathrm{ext}, K)=(28,-0.45)$ mark the parameter values used in calculating the loss function. The model parameters are given in Tab.~\ref{tab_Devalle}. }
\end{figure}

In this network, the macroscopic variables $R(t)$ and $S(t)$ are considered as unavailable for observation. The dependence of the loss function~\eqref{eq:loss_function} on the initial conditions of these variables is eliminated using synchronization methods based on noninvasive and invasive coupling schemes described in Sec.~\ref{sec:methods}. The characteristic value of the feedback gain $K$ for the noninvasive method Eq.~\eqref{eq:model_macr_feed} is estimated by computing the conditional LEs of the mean-field model Eqs.~\eqref{eq:macrDevalle}. Figure~\ref{fig:arnold_lyap_dev}(a) shows the dependence of the maximal conditional LE $\lambda_\mathrm{max}$ on $K$. Synchronization occurs where $\lambda_\mathrm{max}<0$. When calculating the loss function in the DE optimization algorithm, we set $K=0.5$. The characteristic amplitude $K$ and the period $T_\mathrm{ext}$ of the external current in the invasive method are estimated by calculating the synchronization regions (Arnold tongues) in the plane of parameters $(T_\mathrm{ext}, K)$. The regions estimated based on the mean-field Eqs.~\eqref{eq:macrDevalle} with the external current $I_\mathrm{ext}(t)$ defined by Eq.~\eqref{eq:periodicF} are shown in Fig. ~\ref{fig:arnold_lyap_dev}(b).  
The Arnold tongue for inhibitory pulses ($K<0$) is larger than that for excitatory pulses ($K>0$), so we choose the parameters $(T_\mathrm{ext}, K)=(28,-0.45)$ in the inhibitory region to achieve more robust DE optimization.

\subsection{Microscopic and mean-field models of the QIF-AD network}\label{sec:chaos}

Microscopic QIF-AD model describes a heterogeneous population of $j=1,\ldots, N$ neurons that take into account the spike frequency adaptation effects and interacts via a mean-field excitatory coupling. The evolution of the membrane potential $v_j$ and the adaptation variable $a_j$ of a neuron $j$ in a population is governed by the differential equations~\cite{montbrio2025}:
\begin{subequations}
\label{eq:micrMontbrio}
\begin{eqnarray}
\tau_m \dot{v}_j & =& v^2_j+\eta_j+J \tau_m R(t)-a_j+ I_\mathrm{ext}(t), \label{eq:micrMontbrioV}\\
\tau_a \dot{a}_j & =&  -a_j + \beta [\eta_j+J \tau_m R(t)-a_j+ I_\mathrm{ext}(t)]. \label{eq:micrMontbrioA}
\end{eqnarray}
\end{subequations}
Here, as in the previous model, the definition of a QIF neuron requires the resetting rule, i.e., each time a potential $v_j$ reaches infinity, it is reset to minus infinity, and the neuron emits an instantaneous spike that contributes to the mean firing rate $R(t)$, given by the Eq. ~\eqref{eq:R(t)}. $\tau_m$ and $\tau_a$ are the membrane and adaptation time constants, respectively. The heterogeneous currents $\eta_j$ satisfy the Lorentzian distribution Eq.~\eqref{eq:Lorentz}.
The positive parameter $J$ determines the strength of the excitatory coupling, and the positive parameter $\beta$ determines the strength of the adaptation current. $I_\mathrm{ext}(t)$ represents the external uniform current.

In the thermodynamic limit $N \to \infty$, the microscopic Eqs.~\eqref{eq:micrMontbrio} can be reduced to an exact mean-field model defined by a closed system of three differential equations~\cite{montbrio2025} 
\begin{subequations}
\label{eq:macrMontbrio}
\begin{eqnarray}
\tau_m \dot{R} & =& \frac{1}{\pi \tau_m} \frac{\Delta}{1+\beta}+2RV, \label{eq:macrMontbrioR}\\
\tau_m \dot{V} & =& V^2-(\tau_m \pi R)^2 + \bar{\eta}+J \tau_m R - A + I_\mathrm{ext}(t), \label{eq:macrMontbrioV}\\
\tau_a \dot{A} & =&  -A(1+\beta) + \beta [\bar{\eta}+J \tau_m R + I_\mathrm{ext}(t)], \label{eq:macrMontbrioA}
\end{eqnarray}
\end{subequations}
where the definition of the macroscopic variables $R$ and $V$ is the same as in the QIF-IN model, and $A$ is the mean adaptation variable.
This model demonstrates a greater variety of dynamic modes than the QIF-IN model. In addition to periodic collective oscillations, this model can exhibit collective chaotic oscillations. The specific parameter values representing the chaotic oscillations we used in our simulations are given in Tab.~\ref{tab_Mont}.
\begin{table}[h]
\caption{\label{tab_Mont} \textbf{Values of QIF-AD network parameters and optimization bounds used in the DE algorithm.} 
}
\centering
\rowcolors{2}{gray!10}{white} 
\begin{tabular}{cccc}
\rowcolor{gray!30}
Parameters 	& Actual values 	&  Optimization bounds		& Units \\
$\Delta$ 	& 1					& $[0.9,\; 2]$ 		& a.u.\\
$\bar{\eta}$& 3.25				& $[1.75,\; 4.9]$ 	& a.u.	\\
$J$ 			& 20 				& $[10,\; 30]$		& a.u.\\
$\beta$		& 1					& $[0.25,\; 1.25]$  	& a.u.\\
$\tau_m$ 	& 10					& $[7,\; 17]$		& ms \\
$\tau_a$ 	& 100				&  --				& ms 
\end{tabular}
\end{table}

\begin{figure}
\includegraphics[width=0.95\textwidth]{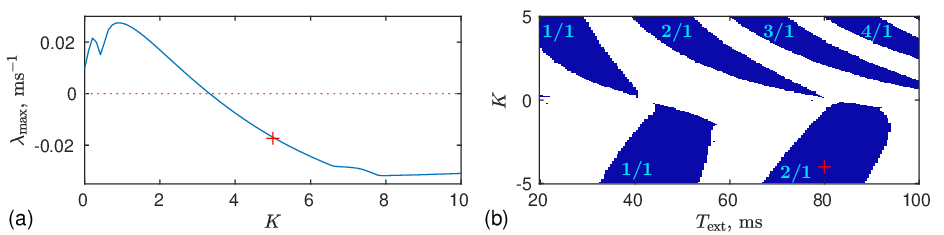}
\centering
\caption{\label{fig:arnold_lyap_mont} Synchronization regions for the QIF-AD network. (a) Maximal conditional Lyapunov exponent as a function of the feedback gain $K$ of the noninvasive method. (b) Arnold tongues in the parameter plane $(T_\mathrm{ext}, K)$ for the invasive method. Arnold tongues are labeled with winding numbers $w=n/1$, where $n$ is the number of spikes generated by the network in one period of the external current. The red crosses in panels (a) $K =5$ and (b) $(T_\mathrm{ext}, K)=(80,-4)$ mark the parameter values used in calculating the loss function. The model parameters are given in Tab.~\ref{tab_Mont}. 
}
\end{figure}

Here, as in the previous model, we assume that only the mean membrane potential is available for observation, while the macroscopic variables $R(t)$ and $A(t)$ are unobservable. To eliminate the dependence of the loss function on the initial conditions of unobserved variables, we again use noninvasive and invasive synchronization methods. Synchronization regions for these methods are shown in Fig.~\ref{fig:arnold_lyap_mont}. The red crosses mark the values of the parameters that we choose to calculate the loss function: $K=5$ for the noninvasive method and $(T_\mathrm{ext}, K)=(80,-4)$ for the invasive method.

\subsection{Numerical integration}

For numerical simulation of microscopic models, it is more convenient to use the equivalent theta neuron formulation by changing variables
\begin{equation}
v_j = \tan(\theta_j/2).
\end{equation}
Then the Eq.~\eqref{eq:micrDevalleV} of the QIF-IN microscopic model transforms to 
\begin{equation}
\tau_m \dot{\theta}_j  =  1-\cos(\theta_j) + \left[1+\cos(\theta_j) \right] \left[\eta_j - J \tau_m S +I_\mathrm{ext}(t) \right].
\end{equation}
Similarly, the Eq.~\eqref{eq:micrMontbrioV} of the QIF-AD microscopic model takes the following form:
\begin{equation}
\tau_m \dot{\theta}_j  =  1-\cos(\theta_j) + \left[1+\cos(\theta_j) \right] \left[\eta_j+J \tau_m R(t)-a_j+ I_\mathrm{ext}(t) \right].
\end{equation}
The theta neuron formulation avoids the discontinuity problem. When the membrane potential $v_j$ of the QIF neuron rises to $+\infty$ and falls to $-\infty$, the phase $\theta_j$ of the theta neuron just crosses the value $\pi$.

The mean firing rate $R(t)$ and mean membrane potential $V(t)$ for both QIF-IN and QIF-AD microscopic models were computed via the conformal mapping of the complex-valued Kuramoto order parameter $Z(t)$~\cite{montbrio2025}:
\begin{equation}
R(t) = \frac{1}{\pi \tau_m}\Re \frac{1-Z^*(t)}{1+Z^*(t)}, \quad V(t) = \Im \left[\frac{1-Z^*(t)}{1+Z^*(t)}\right], \quad Z(t) = \frac{1}{N}\sum_{j=1}^N \exp(i\theta_j)).
\end{equation}
The mean adaptation variable of the QIF-AD microscopic model is defined as $A(t)=\sum_{j=1}^N a_j(t)/N$. The parameter values $\eta_j$ satisfying Lorentzian distribution were generated deterministically by using the formula $\eta_j = \bar{\eta} + \Delta \tan(\pi \left[ \frac{1-2\varepsilon}{N-1} (j-1) - \frac{1}{2}+\varepsilon  \right] )$, where $j=1,\ldots, N$ and $\varepsilon = 10^{-3}$. The small parameter $\varepsilon$ is added to prevent the tangent function from diverging at the points $\pm \pi/2$.

The integration of the microscopic equations in the theta neuron formulation was performed using the 4th-order Runge-Kutta method with a time step of $\delta t = 10^{-2}$ ms for both the QIF-IN and QIF-AD network models. This time step is also interpreted as the sampling step of the network output.

\section*{Data availability}
The generated datasets and  the source code will be available to download after the publication.

\section*{Competing interests}
The authors declare no competing interests.

\section*{Authors contributions}
Conceptualization: K.P. Methodology: K.P, I.R. Numerical experiments: I.R. Data analysis: K.P., I.R. Writing-Review-Editing: K.P., I.R. Funding acquisition and project administrator: K.P. All authors have read and approved the final version of the manuscript for publication.

\section*{Funding}
This work was supported by Grant No. S-MIP-24-57 of the Research Council of Lithuania.

\bibliography{ms_syst_identification}

\end{document}